\documentclass[10pt, letterpaper, conference, compsocconf]{IEEEtran}
\IEEEoverridecommandlockouts
\ifCLASSOPTIONcompsoc
  \usepackage[nocompress]{cite}
\else
  \usepackage{cite}
\fi

\ifCLASSINFOpdf
\else
   \usepackage[dvips]{graphicx}
\fi
\usepackage[cmex10]{amsmath}

\usepackage{url}

\begin{document}
\title{Reinventing the Triangles:\\ Rule of Thumb for Assessing Detectability}

\author{\IEEEauthorblockN{Jeremi K. Ochab}
\IEEEauthorblockA{M. Smoluchowski Institute of Physics\\
Jagiellonian University\\
Cracow, Poland\\
Email:jeremi.ochab@uj.edu.pl
}
}

\maketitle

\begin{abstract}
Based on recent advances regarding eigenspectra of adjacency matrices, derived from the random matrix theory, the paper reintroduces the global clustering coefficient as a simple means of assessing if nonrandom modules can be detected in undirected, unweighted networks.
\end{abstract}


%
\IEEEpeerreviewmaketitle

\section{Introduction}
Across disciplines one of the most relevant features of complex networks is their modular structure. It is crucial in social analysis, it defines or represents the function of biological networks, it affects the processes taking place on networks.
The modules, also called communities, bring information: they allow to visualize and understand data, and to predict behavior.

Consequently, \textit{community detection} is at the forefront of data mining techniques.
Even though numerous\textemdash and at times surprisingly effective\textemdash algorithms of community detection have been appearing now for a decade, there remains a long-standing question of statistical significance of the modules discovered in networks.
Several approaches were already described in \cite{SF_review} (see Sec. 14, \textit{Significance of clustering} therein), but new techniques and theoretical advances, notably those stemming from the random matrix theory, have since been brought into play.

As described in the following section, the key problem with assessing significance is caused by the detectability limit. Our aim is to propose a fast and efficient rule-of-thumb to discriminate between networks in which there can or cannot hide significant communities.

\section{Rationale--Limits of Detectability}
\label{sec:detect}

It has been known that community detection algorithms tend to find some clusters even in random graphs (RG), that they fail to detect clusters in graphs which are not fully random, and that in RG ensembles themselves there are instances of graphs in which communities appear solely by chance. In order to alleviate these problems one may apply bootstrap procedures \cite{AL_Consens}, which significantly enhance the partitioning. Alternatively, the problem has been tackled with the aid of extreme value statistics \cite{AL_Sign}, so that graph partitions can be assigned a statistical significance score.

However, Reichardt and Leone \cite{Leone} have shown that a sharp phase transition exists between undetectable and detectable cluster structures. Such a divide means that below the transition community structure found by an algorithm is unreliable. Further studies by Decelle et al. \cite{Decelle} confirm that the transition exists, in some cases however it is possible\textemdash although exponentially hard\textemdash to find true clusters below this limit.

In a series of papers that followed \cite{MEJN1,MEJN2,MEJN3}  Nadakuditi, et al., have demonstrated a connection between spectra of adjacency matrices of graphs and detectability. Namely, the eigenvalues of the adjacency matrix either belong to a “bulk” (a continuous spectrum) or are isolated; in graphs without a community structure the only isolated eigenvalue is the largest one; any new community introduces a new isolated eigenvalue. Conversely, whenever an eigenvalue associated with a given community hides into the bulk, it becomes undetectable.
From this perspective, the knowledge of a graph's eigenspectrum should be sufficient to answer the question of significance of clusters, and it seems that the spectral methods of community detection ought to be favored. That being said, one issue remains: calculating the spectra is computationally expensive.

Thus, the aim is to quantify “detectability” by means of a single parameter which can be computed fast and easily for any given network. We argue that global clustering coefficient can be applied for that purpose.

\section{Reminder--Clustering Coefficients}

It has been long years since the clustering coefficient has become widely used to benchmark empirical networks and new\textemdash small-world \cite{WS1998} and scale-free networks\textemdash against random Erd\H{o}s-R\'{e}nyi (ER) graphs. Unsurprisingly, one such number cannot describe a real complex network; later studies focused on degree distributions, path lengths, node correlations or quantities describing percolation, epidemic spread, and many others. Although the clustering coefficient(s) gradually ceased to be interesting to measure, databases such as Stanford Large Network Dataset Collection \cite{SNAP} still provide it, along the number of triangles, as a fundamental graph statistic. Indeed, our paper advocates this measure as a simple and fast (especially that often it is precalculated) benchmark for assessing the significance of community structure.

Since there has been some confusion about what precisely is the clustering coefficient, let us define it: by \textit{local clustering coefficient} $C_l$ of node $l$ we mean
\begin{equation}
C_l=N_{l,\triangle}/\binom{k_l}{2},
\end{equation}
where $N_{l,\triangle}$ is the number of triangles formed by the edges $(l,m)$, $(l,n)$ and $(m,n)$, where $m$ and $n$ are nodes adjacent to $l$; $k_l$ is the degree of $l$ (i.e., the number of edges incident to $l$), and the denominator is the number of all pairs of edges incident to node $l$.

The average or \textit{mean local clustering coefficient} often found in the literature is simply
\begin{equation}
C=\frac{1}{N}\sum_{l=1}^{N}C_l,
\end{equation}
where $N$ is the size of the network (i.e., the number of nodes).
As argued by Bollob\'{a}s \cite{Boll}, it ``is  often  not  very  informative", therefore we do not use it further. Rather, we choose to work with what we will refer to as \textit{global clustering coefficient} (GCC), customarily defined as
\begin{equation}
C=\frac{3\times \text{number of triangles}}{\text{number of pairs of adjacent edges}}.
\label{eq:gcc_tr}
\end{equation}
These definitions have been further extended to the cases of directed \cite{Fagiolo} and weighted \cite{Barrat,Onnela} networks.

As mentioned above, these coefficients have been used as a way of comparison with Erd\H{o}s-R\'{e}nyi graphs, for which it is known to be
\begin{equation}
C_{ER}=\frac{\langle k \rangle}{N-1},
\end{equation}
where $\langle \cdot \rangle$ denotes averaging, and $\langle k \rangle$ is simply the average degree.
For uncorrelated graphs with a given degree distribution the formula \cite{MEJN_clust,Boguna_clust,ZB_clust,Dorog_clust}
\begin{equation}
C_{\mathrm{uc}}\approx \frac{1}{N}\frac{(\langle k^2 \rangle-\langle k \rangle)^2}{\langle k \rangle^3}
\label{eq:uc}
\end{equation}
is widely used.

As an example of a scale-free network, Bollob\'{a}s \cite{Boll} derives and proves a corrected (after \cite{Klemm}) estimate on the clustering coefficient for Barab\'{a}si-Albert (BA) networks
\begin{equation}
C_{BA}=\frac{m-1}{8}\frac{(\log N)^2}{N},
\label{eq:BA}
\end{equation}
where $m$ is the number of edges that a new node appears with in the process of network growth.
We use BA graphs in Fig. \ref{fig:over} as an example of unclustered scale-free networks to show behavior of various estimates of $C$.

\section{Observations}
\subsection{GCC vs Mixing Parameter}
\label{sec:gcc_mix}
\begin{figure}[!t]
\centering
\includegraphics[width=2.5in]{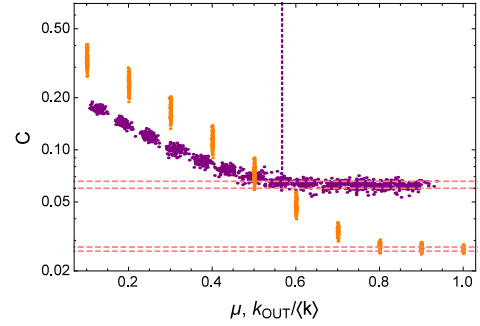}
\caption{Log plot of GCC versus the mixing parameter ($\mu$ for LFR, and $k_{OUT}/\langle k \rangle$ for NG networks; the ensembles are described in text). Dashed horizontal lines indicate mean $\pm$ standard deviation of GCC for respective samples of random graphs. The dotted vertical line shows the theoretical detectability limit for NG graphs as given by \cite{MEJN1}. The undetectability regime overlaps with the region where GCC curves saturate.}
\label{fig:corrPlot}
\end{figure}

The critical observation is that for ensembles of benchmark graphs with tunable strength of community structure the theoretically predicted limit of detectability coincides with the region where GCC saturates, as shown in Fig.~\ref{fig:corrPlot}.

Each of the points in the plot represents a single graph. There are two benchmark ensembles used: Newman-Girvan (NG) \cite{NG_Bench}, whose graphs consist of 4 equally-sized communities (ER type with the probability of intercommunity linking a pair of nodes $p_{\text{out}}$ and intracommunity linking probability $p_{\text{in}}$), mean degree $\langle k \rangle= 16$, and $N=256$ nodes; Lancichinetti-Fortunato-Radicchi (LFR) \cite{AL_Bench}, with $\langle k \rangle=20$, maximal degree $k_{\text{MAX}}=50$, power-law degree distribution with exponent $\gamma=2$, power-law community size distribution with exponent $\gamma_c=1$, allowed range of community sizes $20-100$ nodes, and the network size $N=1000$ (we retain the parameters used originally in \cite{AL_Compar} for the sake of comparison).
In the case of LFR graphs, the parameter quantifying community strength is the mixing parameter $\mu$, which is a fraction of links of a node pointing outside of its community.
In the case of NG graphs, this role is taken by $k_{OUT}/\langle k \rangle$, where $k_{OUT}$ is the number of links pointing outside of a node's community; it is simpler than $\mu$, since the degree distribution is well peaked around its mean value. For each mixing parameter value, 100 networks were sampled (clouds of points visible in the plot).

For these ensembles, the fully random connections form for $k_{OUT}/\langle k \rangle \approx 12/16$ for NG and $\mu=1$ for LFR. In both cases, community detection algorithms fail much earlier\cite{AL_Compar}. The theoretical position of detectability limit as calculated in \cite{MEJN1} is in this case $k_{OUT}/\langle k \rangle \approx 9/16$; unfortunately, the calculations for LFR benchmark are much harder to obtain. 

The reason why it was possible to find the detectability limit theoretically, is the fact that we already know the structure of NG networks. In the limit of large network size and, what is more important, large average degree, Nadakuditi et al. have shown how to calculate the eigenspectra of adjacency matrices of stochastic block model \cite{MEJN1}, configuration model \cite{MEJN2}, and a generalization of the two \cite{MEJN3}, which allows to find the detectability limits for networks with arbitrary degree distributions and community structure. Nevertheless, in practice we do not know the structure in the first place, hence we do not know the full eigenspectrum, and so we cannot decide whether a given network has any communities within detectability regime or not. Now, let us try to deduce that from the value of GCC.

\subsection{GCC vs Eigenspectrum}

The reason that the global clustering coefficient can be an index of detectability is fairly straightforward. Let us rewrite (\ref{eq:gcc_tr}) as follows
\begin{equation}
\label{eq:gcc}
C=\frac{\sum_{l=1}^{N}N_{l,\triangle}/N}{(\langle k^2 \rangle -\langle k\rangle)/2}
=\frac{\sum_{l=1}^{N}A^3_{ll}/N}{\langle k^2 \rangle -\langle k\rangle}
=\frac{\sum_{l=1}^{N}\lambda_l^3/N}{\langle k^2 \rangle -\langle k\rangle},
\end{equation}
where $A_{ij}$ is an element of the adjacency matrix $\mathbf{A}$, $\lambda_i$ is its eigenvalue, and where we have used $\sum_{l=1}^{N}\binom{k_l}{2}= N(\langle k^2 \rangle -\langle k\rangle)/2$ and $N_{l,\triangle}=A^3_{ll}/2$. The second equality in (\ref{eq:gcc})comes from the fact that $(\mathbf{A}^n)_{ij}$ encodes the number of paths of length $n$ between nodes $i$ and $j$; the third equality comes from the fact that thanks to the cyclic property of the trace of a matrix, it is invariant to orthogonal transformations.

Now, the simple fact that clustering coefficients are connected to the third moment of the eigenspectrum was known \cite{Barabasi}. Yet, as explained in Sec. \ref{sec:detect}, in undirected and unweighted networks communities appear as isolated eigenvalues to the right of the bulk of the spectrum. This means that the spectrum becomes more skewed to the right; in other words, the third moment of the spectrum\textemdash and consequently GCC\textemdash increases. Such a behavior is consistent with the observation in Sec. \ref{sec:gcc_mix}.

\subsection{Estimating Baseline GCC}

The forthcoming issue is how to calculate the reference value of GCC, i.e., the value for a network without communities. One straightforward approach is to randomize the network under study, while keeping constant other of its characteristics that can affect GCC, mainly the degree distribution (or even degree sequence). Notwithstanding, this method is computationally costly, since a whole ensemble of randomized graphs is needed to make the comparison with the original empirical network meaningful.

Another approach is to use a theory to reproduce the eigenspectrum of graphs with a given degree distribution, calculate its third moment, and obtain GCC. Indeed, it can be accomplished, e.g., following \cite{MEJN3}.
Under assumptions on the community structure, e.g., as obtained from a community detection algorithm, the isolated eigenvalues can also be separately calculated.
For large scale-free graphs, however, solving numerically the resulting equation for a specified degree sequence can become increasingly infeasible, as the number of terms (and solutions) also increases with the number of distinct degrees in the network.

\begin{figure}[!t]
\centering
\includegraphics[width=2.5in]{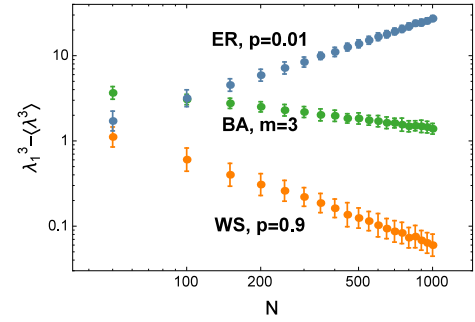}
\caption{Log-log plot of minus third moment of the bulk of eigenspectrum for Barab\'{a}si-Albert, Erd\H{o}s-R\'{e}nyi, and Watts-Strogatz networks with the given 
parameters. Error bars are standard deviations for samples of 100 graphs each.}
\label{fig:skew}
\end{figure}

More importantly, and to our demise, the bulk of the spectrum thus obtained is strictly symmetric. Consequently, the whole third spectral moment of a random graph will come from the largest eigenvalue. Unfortunately, we observed that for ER, BA networks, and even Watts-Strogatz \cite{WS1998} networks with sufficiently high rewiring probability the third moment of the bulk is negative\footnote{It should be noted that in graphs which are bipartite, or even partly so\textemdash e.g., WS graphs with small $p$\textemdash negative isolated eigenvalues can appear and thwart our efforts. Hence, we assume only positive isolated eigenvalues.}, see Fig.  \ref{fig:skew}. 
Consequently, for a wide range of real-world networks we can expect
\begin{equation}
C < \frac{1}{N}\frac{\lambda_1^3}{\langle k^2 \rangle -\langle k\rangle}
\label{eq:gcc<}
\end{equation}
to hold. (This inequality can also be seen as a lower bound for $\lambda_1 > \sqrt[3]{\text{number of triangles}}$ which is usually stronger than the known bound $\lambda_1>\sqrt[3]{2\max_{l}N_{l,\triangle}}$.)

\begin{figure}[!t]
\centering
\includegraphics[width=2.5in]{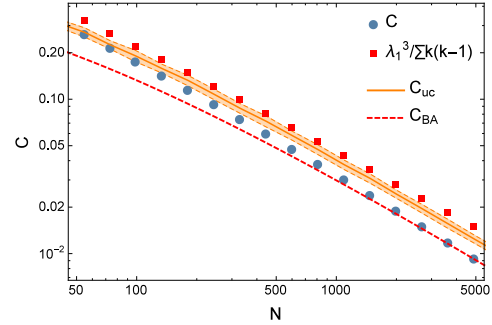}
\caption{Comparison of global clustering coefficients $C$ (average over a sample of 100 BA graphs; standard deviations are of the size of the points), theoretical predictions obtained for uncorrelated networks with a given degree distribution $C_{\mathrm{uc}}$ (\ref{eq:uc}) (shaded orange region), for BA (\ref{eq:BA}) (red dashed line), and the contribution of the largest eigenvalue to GCC (red squares). When the network structure is unknown, the predictions are overestimated by more than the sample error of GCC.}
\label{fig:over}
\end{figure}

It follows that the theoretical prediction on GCC as calculated using \cite{MEJN3} is largely overestimated (i.e., an order of magnitude larger than the standard deviation of an ensemble with a given degree distribution). An example of such a behavior is shown for BA networks in Fig.\ref{fig:over}. The only theoretical prediction that could be useful for our purposes is (\ref{eq:BA}); it is, however, a prediction specifically derived for this particular type of network. When analyzing real-world graphs, we lack such conveniences.

\section{Results}

At this point we ought to recapitulate our aims, collect the pieces of information we have, and formulate the assumptions under which we will be able to proceed.

The aim is: based on the global clustering coefficient (\ref{eq:gcc_tr}) to construct a criterion that decides whether
\begin{enumerate}
\item	a given network has a significant community structure,\label{aim1}
\item	a given partition of the network is statistically significant,\label{aim2}
\item	a given community of the partition is statistically significant.
\end{enumerate}
In this paper we will propose an answer to \ref{aim1}) and validate it,
as well as suggest a way to develop a criterion for \ref{aim2}).

Let us list the (empirical) facts we have gathered up to this point:
\begin{enumerate}
\item	GCC is proportional to the third moment of the eigenspectrum,
\item	(empirical) for graphs with a community structure GCC is greater than for random graphs, but it reaches the RG value before its connectivity is fully random, see Fig.~\ref{fig:corrPlot},
\item	(empirical) GCC is strictly lower than what can be predicted by the largest eigenvalue alone, see (\ref{eq:gcc<}),\label{fact}
\item	GCC for the uncorrelated graphs can be approximated by (\ref{eq:uc}).
\end{enumerate}

After GCC, the degree distribution and its two moments, $\langle k \rangle$ and $\langle k^2 \rangle$, have been calculated,
which is the only computational burden, we may proceed further.
One has to keep in mind, nevertheless, following assumptions:
\begin{enumerate}
\item	the network is not bipartite, see Footnote 1., 
\item	the network is uncorrelated, so that (\ref{eq:uc}) is legitimate,
\item	$\lambda_1 \geq \langle k^2 \rangle/\langle k \rangle - 1$,\label{assum}
\item	$N$ is large, as always.
\end{enumerate}
Let us note that for the Poisson degree distribution the Assumption \ref{assum}) is always true, since $\lambda_1 \geq \sqrt{\langle k^2 \rangle}=\sqrt{\langle k \rangle^2+\langle k \rangle} > \langle k \rangle = \langle k^2 \rangle/\langle k \rangle -1$, which comes from Rayleigh's inequalities (see e.g. \cite{Mieghem}). For scale-free networks, however, the second inequality is inverted and there may appear graphs breaking Assum. \ref{assum}). (We did find such instances for BA networks but, surprisingly, not for LFR which are scale-free as well.)

\subsection{Procedure for an unpartitioned graph}

In the case of Aim \ref{aim1}), the way to proceed is embarrassingly simple: if
\begin{equation}
C \leq C_{\mathrm{uc}}
\label{eq:cond1}
\end{equation}
is true, then the network is outside the detectability regime, i.e., there are no significant communities.

A remark is necessary: let us notice that the above equation is a conservative criterion.
The condition that we should have checked is in fact (\ref{eq:gcc<}). But it can be easily checked that the approximation $C_{\mathrm{uc}}$ corresponds to $\lambda_1 \approx \langle k^2 \rangle/\langle k \rangle-1$.

However, since the true largest eigenvalue is unknown, we can utilize the well-known approximation by Chung et al.\cite{Chung},
\begin{equation}
\lambda_1 = \left(1+o(1)\right)\max \left\{ \frac{\langle k^2 \rangle}{\langle k \rangle} , \sqrt{k_{\mathrm{MAX}}} \right\},
\label{eq:Chung}
\end{equation}
which in most cases we are interested in yields $\langle k^2 \rangle/\langle k \rangle$. Consequently, the criterion (\ref{eq:gcc<}) should be given by 
\begin{equation}
C < \frac{1}{N}\frac{\langle k^2 \rangle^3}{\langle k \rangle^3(\langle k^2 \rangle -\langle k\rangle)}.
\label{eq:cond2}
\end{equation}
By now, it should be clear why the Assumption \ref{assum}) was listed before: as long as it is fulfilled the criterion given in (\ref{eq:cond1}) is stronger than (\ref{eq:gcc<}).

Though (\ref{eq:Chung}) is proved to be the upper bound of $\lambda_1$, in Fig.~\ref{fig:over} the approximation would be even higher than the one resulting from true value of $\lambda_1$ (red squares). This means that in some cases the factor $o(1)$ is of a magnitude which can affect our inference.
Finally, the ordering of the theoretical predictions, as depicted in Fig. \ref{fig:over}, will be of use to us.\\

Even though, condition (\ref{eq:cond1}) does not provide us with any error estimates, we still believe that the approximate position of $\lambda_1$ is given by (\ref{eq:Chung}).
As a result we can use (\ref{eq:cond2}) as an upper bound for an error estimate.
Thus, the rule-of-thumb is as follows:
\begin{itemize}
\item if condition (\ref{eq:cond1}) is true, there is no detectable community structure,
\item if condition (\ref{eq:cond1}) is false, but (\ref{eq:cond2}) is true, either there is no detectable community structure or there are some detectable and some undetectable communities,
\item if both are false, there is some detectable community structure.
\end{itemize}

\begin{figure}[!t]
\centering
\includegraphics[width=2.5in]{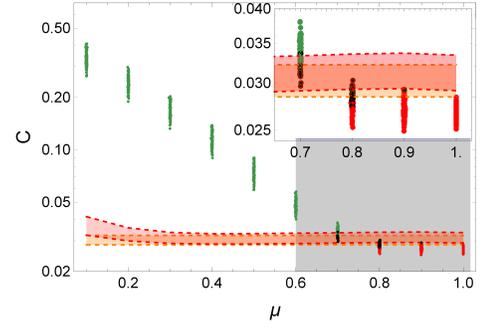}
\caption{Log plot of GCC for LFR $N=1000, B$ benchmark, as in Fig. \ref{fig:corrPlot}. The red shaded strip shows the contribution of true $\lambda_1$ to $C$; the orange shaded strip shows the region between conditions (\ref{eq:cond1}) and (\ref{eq:cond2}). The red dots represent graphs fulfilling (\ref{eq:cond1}); the black dots fulfill (\ref{eq:cond2}), but not (\ref{eq:cond1}); the green dots fulfill neither. The grey region indicates where all known community detection methods gradually begin to fail. The inset magnifies the region of (un)detectability transition.}
\label{fig:criteria}
\end{figure}

To validate above criteria we again resort to LFR benchmark graphs, since all of the state of the art community detection methods have been tested on them, including Infomap \cite{Infomap}, the Louvain method \cite{Louvain} or Oslom \cite{Oslom}.
Tests on these graphs, with exactly the same parameters, have shown \cite{AL_Compar,AL_Consens} that all known methods begin to decrease their accuracy at $\mu \approx 0.6$, and fail after crossing $\mu \approx 0.7-0.8$.
As illustrated in Fig.~\ref{fig:criteria}, this is in accord with the criteria given above.

To address the Aim \ref{aim2}) above, let us assume that
we already have obtained a partition of the graph into $q$ clusters.
For instance, it might be a result of any of the algorithms we have mentioned,
or it might be the planted partition, e.g., in LFR benchmarks
(note that by construction these graphs have a partition planted even for $\mu=1$).
Following \cite{MEJN3}, it is possible to calculate the $q$ discrete eigenvalues
that are theoretically predicted to appear in such a partitioned graph.
We are then able to compute their contribution to $C$ in (\ref{eq:gcc_tr}),
just as we did for $\lambda_1$, and accordingly revise the criteria (\ref{eq:cond1})-(\ref{eq:cond2}).
Note, that instead of using approximation (\ref{eq:Chung}) for $\lambda_1$ in this Section,
we may use \cite{MEJN3} for the largest eigenvalue only.
Depending on the computational resources one has,
it might be feasible to simply calculate $\lambda_1$ numerically\textemdash
although the theoretical approximations seem sufficient,
an improvement on the error estimate of roughly $20\%$ can be expected (at least on graphs akin to LFR),
as can be seen in the inset of Fig.~\ref{fig:criteria}.


\section{Conclusion}

We have proposed a simple method to deduce whether it is possible to detect communities in a given network.
Since the only computations needed are: calculating triangles and the degree distribution, we claim this method is fast and computationally inexpensive.
The state of the art algorithm for an exact triangle enumeration runs in $O(E^{1.41})$, where $E$ is the number of edges in the graph. Faster approximate algorithms do exist \cite{triangles}, whose accuracy of around $95\%$ are sufficient, although introduce an additional error. Although there exist algorithms of community detection linear in time, assessing statistical significance of the partitions obtained usually involves bootstrapping which is not favorable.

The error bound in procedure we suggested is rather heuristic;
however, the approximations of $\lambda_1$ (\ref{eq:Chung}) and GCC (\ref{eq:uc}) can be improved, e.g., if we have ansatz on degree-degree correlations, which in some cases are known \cite{Redner_BA}. Still, if the assumption \ref{assum}) holds, the criteria are conservative and on the safe side. 

The natural extension of the scheme presented here is: directed and weighted networks, for which generalized versions of clustering coefficients have been proposed \cite{Fagiolo,Barrat,Onnela}.
Initial computations on directed LFR benchmarks show that analogous criteria are plausible. For weighted networks, however, the clustering coefficients behave nonmonotonously as a function of the mixing parameter, which introduces additional complications.

Further systematic comparison with other methods is needed and,
more importantly, attacking real-world networks.


\ifCLASSOPTIONcompsoc
  \section*{Acknowledgments}
\else
  \section*{Acknowledgment}
\fi

My thanks go to Santo Fortunato for the discussions at the early stage of the study and a week's stay at BECS, Aaalto University and to Zdzis\l{}aw Burda for consultations throughout. I also thank an Anonymous Reviewer for useful insights about bounds for the principal eigenvalue.
The work has been funded by Grant No. DEC-2013/09/N/ST6/01419 of the National Science Centre of Poland.


\end{document}